\documentclass{article}

\usepackage[margin=1.25in]{geometry}
\usepackage[english]{babel}
\usepackage[final]{graphicx}
\usepackage{bm}
\usepackage{amsmath}
\usepackage[square,sort&compress,numbers]{natbib}
\usepackage[final,hidelinks]{hyperref}
\usepackage[capitalise]{cleveref}
\usepackage[format=plain,labelfont={bf},textfont=it]{caption}

\newcommand{\ea}{\textit{et al.}}
\newcommand{\ie}{\textit{i.e.}}
\newcommand{\eg}{\textit{e.g.}}

\newcommand{\bs}[2]{\mbox{#1\hspace{0.1em}\raisebox{0.18ex}{--}\hspace{0.1em}}#2}
\newcommand{\bd}[2]{\mbox{#1\hspace{0.3em}\raisebox{0.18ex}{=\hspace{-0.8em =}}\hspace{0.3em}}#2}

\newcommand{\TWrho}{$\rho$}
\newcommand{\TWgrad}{\mbox{$\bm{\nabla}$\hspace{-1.2pt}$\bm{\rho}$}}
\newcommand{\TWhess}{$H$\hspace{-0.13em}\TWrho}
\newcommand{\TWlap}{$\nabla^2$\hspace{-0.6pt}\TWrho}

\newcommand{\TWcgrad}{$\bm{\nabla}$\hspace{-1.2pt}$\cal{\bm{P}}$}
\newcommand{\TWt}{${\cal T}$}

\newcommand{\TWp}{${\cal P}$}
\newcommand{\TWpi}{${\cal P}_i(\theta,\phi)$}

\newcommand{\TWdgb}{$d$GB}
\newcommand{\TWdgbi}{$d{\rm{GB}}_i(\theta,\phi)$}

\newcommand{\TWcgp}{${\cal{G}}$}
	
\begin{document}

\title{The Quantum Theory of Atoms in Molecules in Condensed Charge Density Space}

\author{Timothy R.~Wilson\thanks{Molecular Theory Group, Colorado School of Mines, 1500 Illinois St., Golden, Colorado, USA} \and M.~E.~Eberhart\footnotemark[1] \thanks{\textit{Corresponding author:} \href{mailto:meberhar@mines.edu?subject=Quantum\%20Theory\%20of\%20Atoms\%20in\%20Molecules\%20in\%20Condensed\%20Charge\%20Density\%20Space\%20(Arxiv.org)}{meberhar@mines.edu}}}

\maketitle

\begin{abstract}
	By leveraging the fundamental doctrine of The Quantum Theory of Atoms in Molecules---the partitioning of the electron charge density (\TWrho) into regions bounded by surfaces of zero flux---we map the gradient field of \TWrho\ onto a 2D space called the gradient bundle condensed charge density (\TWp).
	The topology of \TWp\ appears to correlate with regions of chemical significance in \TWrho.
	The bond wedge is defined as the image in \TWrho\ of the basin of attraction in \TWp, analogous to the Bader atom, which is the basin of attraction in \TWrho.
	A bond bundle is defined as the union of bond wedges that share interatomic surfaces.
	We show that maxima in \TWp\ typically map to bond paths in \TWrho, though this is not necessarily always true.
	This observation addresses many of the concerns regarding the chemical significance of bond critical points and bond paths in The Quantum Theory of Atoms in Molecules.
\end{abstract}

\section{Introduction}
\label{sec:Intro}

Richard Bader's Quantum Theory of Atoms in Molecules (QTAIM) \cite{bader1994} has been recognized as the only fully quantum mechanics-based theory that enables the application of traditional chemical concepts in an unambiguous manner \cite{shahbazian2006}.
This achievement stems from the decomposition of a molecule or solid's electron charge density (\TWrho) into non-overlapping proper open subsystems, each bound by a surface over which the flux in its gradient (\TWgrad) vanishes \cite{bader1994,nasertayoob2008}.
It was assumed that these subsystems possessed a well-defined energy, though recently, Anderson \ea\ have questioned this assumption \cite{anderson2010}.

While there are an infinite number of regions over which this zero flux condition is satisfied---regions now referred to as quantum divided basins \cite{heidarzadeh2011}---Bader realized that every nucleus is bounded by a unique surface over which the flux of \TWgrad\ is everywhere zero, denoted as a zero flux surface.
These regions Bader recognized as the atoms in molecules, which he called atomic basins, though they are also referred to as Bader atoms.
Each of these atoms is characterized by a nonarbitray boundary, an unambiguous electron count, and, mindful of the concerns expressed in ref.~\citenum{anderson2010}, a well-defined energy.

With a single insight Bader provided a theoretically defensible framework from which one can assess and compare atomic properties like size and energy between atomic systems.
Using this framework and within the accuracy of computational or experimental methods, two researchers must arrive at the same conclusion when scrutinizing such properties.
Bader became the leading advocate for the school of thought seeking to reframe all of chemistry in terms of measurable quantities and his belief was that QTAIM would be the central framework from which this new chemical perspective would evolve.

The Bader atom is easily represented in terms of \TWgrad\ and its corresponding critical points (CPs)---the maxima, minima and saddle points where \TWgrad\ vanishes.
An arbitrary gradient path (G) originates from a minimum---called a cage CP, which may be located at infinity---and terminates at a maximum---called a nuclear CP because it is typically coincident with an atomic nucleus.
Equivalent to the zero flux surface-based definition, a Bader atom is the union of all Gs with a shared terminal nuclear CP.
Bader noted that when two atoms share a polyhedral face, their nuclei are connected by a unique G he called a bond path; a ridge along which \TWrho\ is a maximum with respect to all neighboring paths.
Such a path also necessitates the existence of a saddle point of index $-1$ located between bound nuclei and called a bond CP.

Perhaps due to the word ``bond'' in their designations, in our opinion an unwarranted amount of attention has been focused on bond paths and bond CPs, as researchers have repeatedly discovered bond CPs between atoms whose interactions are assumed to be destabilizing \cite{poater2006,haaland2004,keyvani2016}.
These researchers have argued that such points cannot reflect bonding, and recently it has been proposed that ``bond'' should be stricken from their discussion \cite{shahbazian2018}.
However, these discussions are antithetical to the premise of QTAIM, because neither a bond CP nor a bond path constitutes a volume bounded by a zero flux surface, hence they do not have well defined energies, and taken singularly are not required to provide stability information.
Still, the shear number and often extensive analyses of CPs, particularly bond CPs and bond paths, serve to obscure the main focus and strength of QTAIM; the partitioning of a molecule into proper open subsystems.

The Bader atom was recognized as the sole proper subsystem of a molecule until the introduction of the repulsive basin of Pendas \ea\ in 1997 (and later Popelier's \emph{cage} in 2000) \cite{martinpendas1997,popelier2000}.
They argued that just as there are nuclear CP-centered basins there must be cage CP-centered basins satisfying the zero flux condition.
Repulsive basins were defined as the union of gradient paths that originate at a common cage CP.
The zero flux surfaces bounding repulsive basins must contain a number of nuclear CPs.
Bader had originally discounted the significance of such zero flux surfaces due to the cusp in \TWrho\ at nuclear CPs resulting in undefined points in \TWgrad.
However, the nuclear cusp is not real but rather a manifestation of the coulomb approximation to the Schr\"odinger equation.

The absence of a nuclear cusp opened the door to an infinity of zero flux surface-bounded volumes.
Eberhart \cite{eberhart2001} and later Jones and Eberhart \cite{jones2009,jones2010}, introduced one such volume they called the irreducible bundle; a tetrahedral volume that is the simplex of \TWrho, incorporating all four types of CPs---bond, cage, nuclear and ring (a saddle point of index $+1$).
The vertices, edges, and faces of an irreducible bundle are respectively the four types of CPs, the six shortest-length Gs connecting them pairwise, and the four least-area zero flux surfaces with triplets of CPs as their corners.
Like Bader atoms, irreducible bundles share vertices, edges and faces so as to fill space.
The union of irreducible bundles sharing a common nuclear CP gives rise to the Bader atom.
The union of irreducible bundles sharing a common cage CP gives rise to the repulsive basin.
And the union of irreducible bundles sharing a common bond CP gives rise to the bond bundle---a partitioning of the charge density into unique zero flux surface-bounded volumes, each of which contains a single bond path and bond CP.
Bond bundles recover traditional bond properties like bond order \cite{jones2010,goss2018}, and address issues like spurious bond CPs which have been shown to have tiny bond orders \cite{miorelli2015}.

However, unlike Bader atoms and repulsive basins, which are readily apparent from an inspection of \TWgrad, all the boundaries of irreducible bundles are not obvious and are often difficult to locate.
In correspondence with Eberhart [R.~Bader, personal communication, 2004], Bader questioned whether his theory---with only one zero flux surface and a bond path---would lose its elegance due to extensions such as the irreducible bundle, and that with boundaries defined as least area surfaces, QTAIM would become and be perceived as \textit{ad hoc}, where such definitions arise as but a means to specify some unique boundary between bonds.
His point was, at least philosophically, well taken.
And though Bader has passed away, through our deep respect for him we posthumously answer his question; an answer that points to a possible future  research emphasis for QTAIM.

\section{Condensed Charge Density Space}
\label{sec:condencedRhoSpace}

Developing a more satisfying means of identifying the special boundaries of QTAIM involves constructing the space of all volumes bounded by zero flux surfaces, what we call the gradient bundle condensed charge density (\TWp).
Specifically, we will map Gs of \TWrho\ to points in \TWp\ and show that the bond bundle is the topological analogue in \TWp\ of the Bader atom in \TWrho.

\begin{figure}
	\centering
	\includegraphics[width=0.4\textwidth]{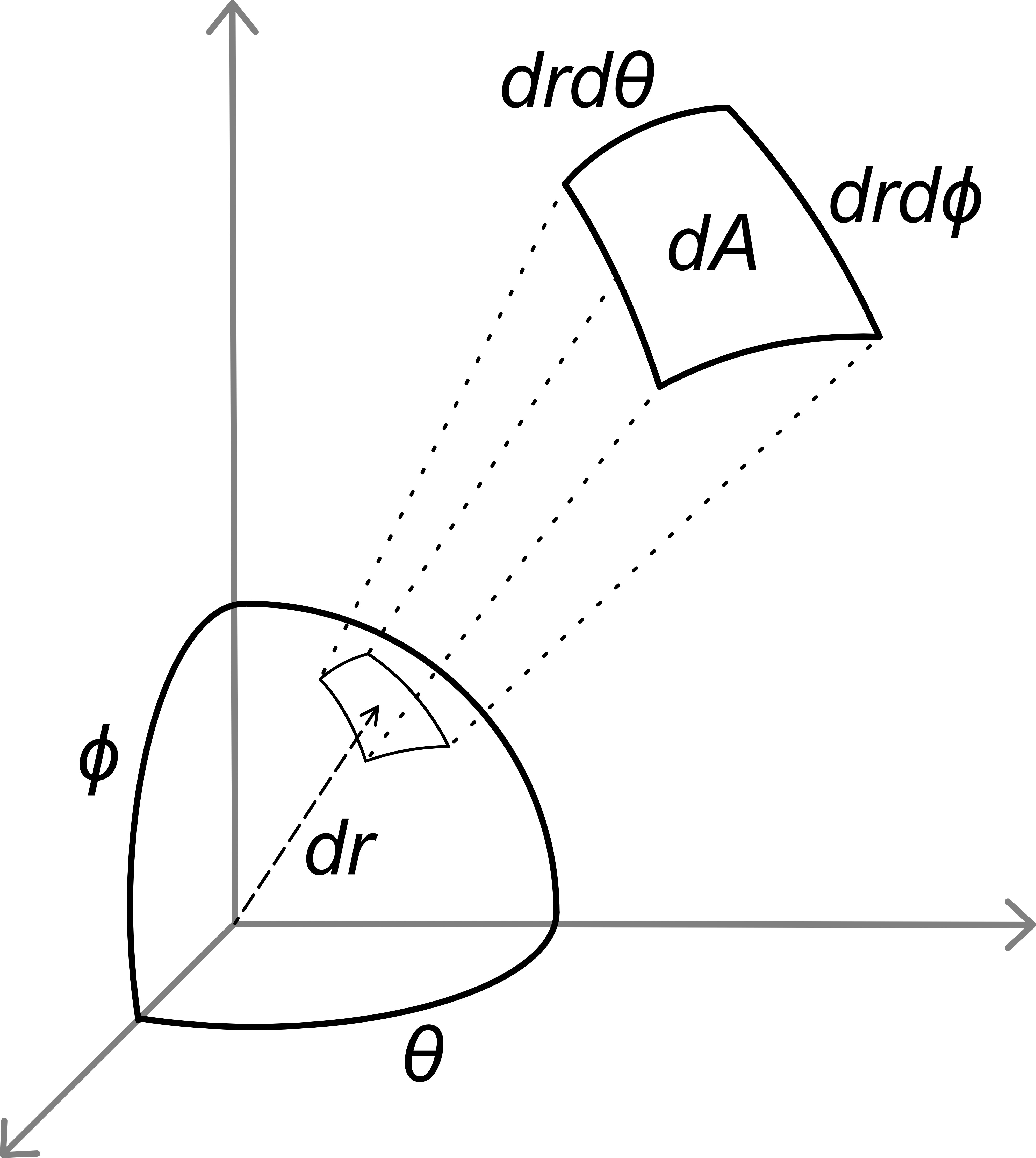}%
	\caption{Differential area element on a sphere.}
	\label{fig:dA}
\end{figure}

\begin{figure}
	\centering
	\includegraphics[width=0.4\textwidth]{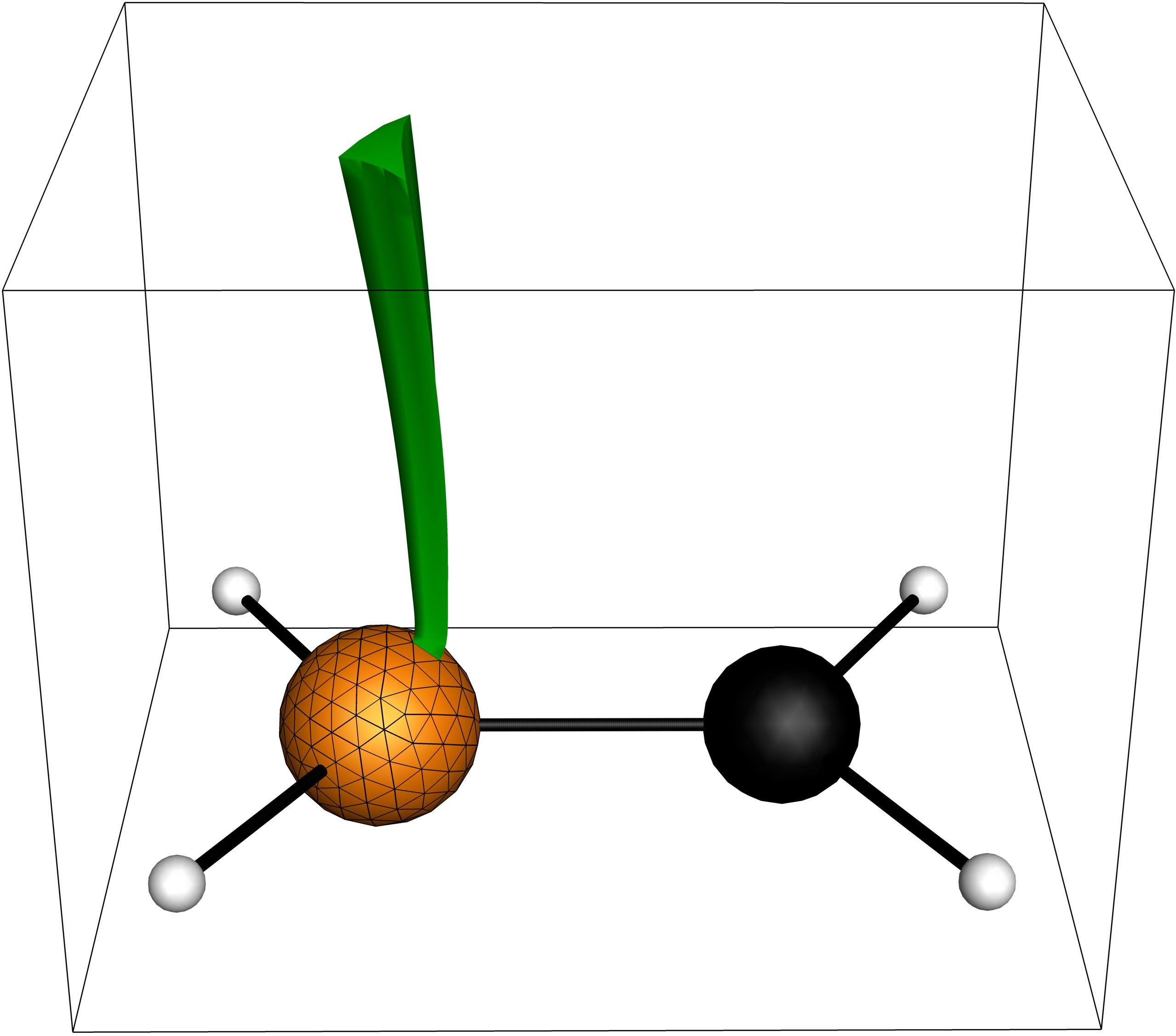}
	\caption{A differential gradient bundle made from gradient paths seeded from the nodes and along the edges of an element on a triangulated sphere.}
	\label{fig:dgb}
\end{figure}

Recall that every G in \TWrho\ originates from a cage CP and terminates at a nuclear CP.
Sufficiently close to its terminus, Gs are radial, making it conceptually convenient to imagine every nuclear CP as the center of a sphere S$_i$ of radius $dr$ that lies within the radial region---in practice, $dr\approx 0.2$\AA\ and would never extend beyond the interatomic surface.
Passing through every point on the surface of these spheres is a G.
The points on a sphere may be specified by a polar and an azimuthal angle, so each of the molecule's Gs may be specified by a pair of coordinates and the index of the nuclear CP at its terminus, \ie\ G$_i(\theta,\phi)$.
Imagine covering each S$_i$ with a set of non-intersecting differential elements of area $dA = d\theta d\phi dr^2$ (\cref{fig:dA}).
The Gs passing through the points interior to each of these area elements gives rise to a family of differential volume elements whose cross sectional area---equal to $d\theta d\phi dr^2$ at S$_i$---changes down their length according to \TWgrad.
These differential gradient bundles, \TWdgbi\ \cite{eberhart2013,morgenstern2015} (\cref{fig:dgb}), 
are the smallest structures bounded by zero flux surfaces and that accordingly possess well-defined energies.
The union of all \TWdgb$_i$ is equivalent to the union of all Gs terminating at nuclear CP $i$ and hence recovers Bader's atomic basin.

\begin{figure}
	\centering
	\includegraphics[width=\textwidth]{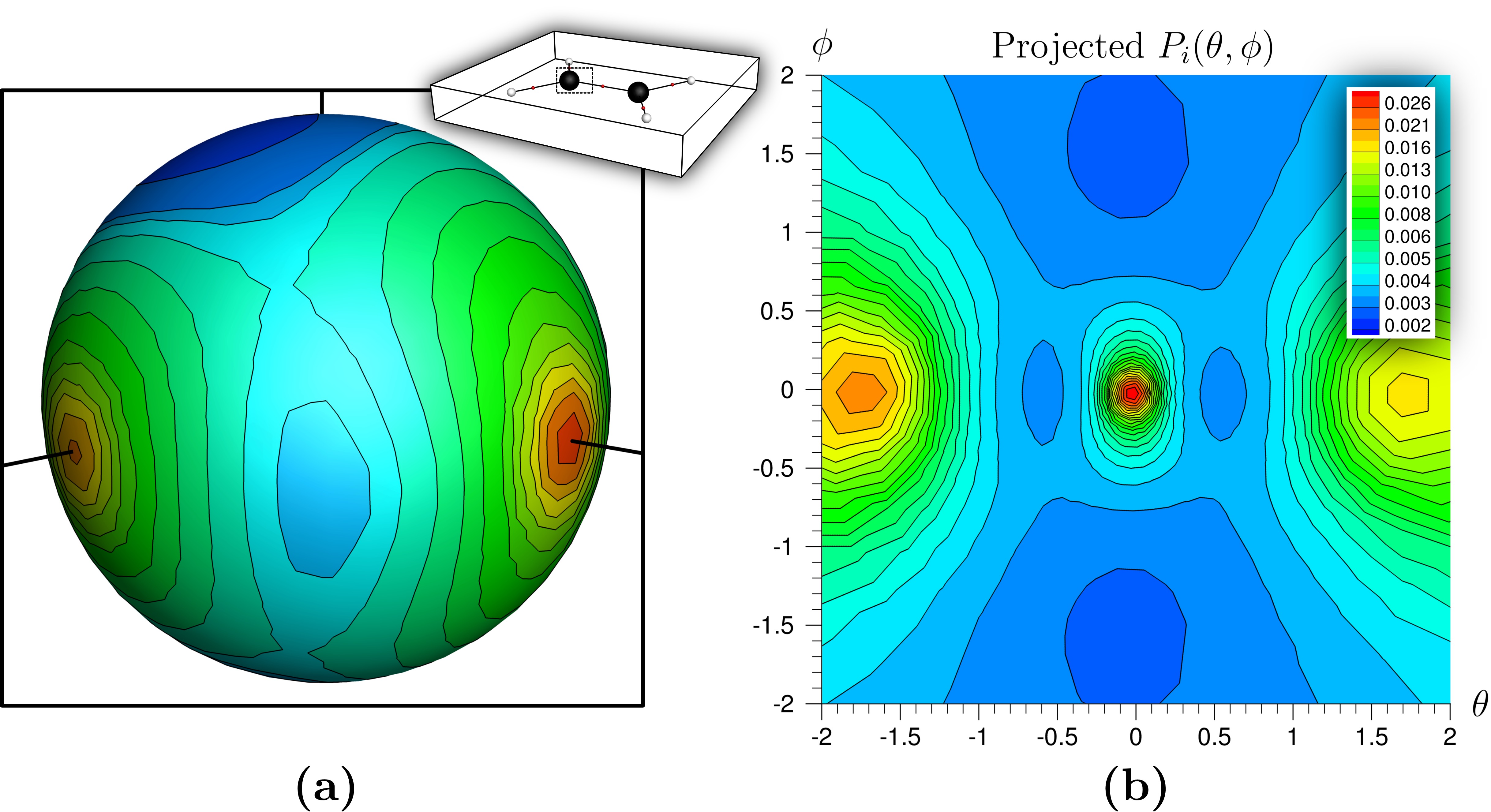}
	\caption{\TWp\ for a carbon atom in ethene.
		Contour band coloring is that of a heat map throughout this paper when a legend is not present, where red and blue indicate high and low values respectively.
		\textbf{a)} Spherical mapping.
		Inset: Black, white, and red spheres respectively indicate carbon nuclear, hydrogen nuclear, and bond CPs (same scheme used when appropriate in remaining figures).
		\textbf{b)} Stereographic projection with C-C bond path at the origin.
		Axes are in units of radians, corresponding to rotation around the sphere in (a).
		\textit{See electronic version for color images.}}
	\label{fig:etheneSphereProjection}
\end{figure}

We define the gradient bundle condensed charge density, \TWpi,\footnote{Not to be confused with the condensed fukui functions} as the area-normalized electron density contained in each \TWdgbi, such that 
\begin{equation*}
	\text{\TWp}_i(\theta,\phi)=\frac{1}{dA(dr)}\int_{\text{G}_i(\theta,\phi)}\rho(s) dA(s)ds,
\end{equation*}
a line integral along $\text{G}_i(\theta,\phi)$ where $s$ is arc length.
\TWp\ is a scalar field with units of electrons per unit area, and maps \TWrho\ within a Bader atom onto a two-dimensional closed surface.
As a visualization tool, this allows one to view what we will show to be the significant features of \TWrho\ at a glance, as in \cref{fig:etheneSphereProjection}a, for a carbon atom in an ethene molecule.

Borrowing terminology from differential geometry, each \TWp$_i$ is referred to as an \textit{atomic chart} and the set of all atomic charts comprising a molecule is termed its \textit{molecular atlas}.
As an alternative to representing atomic charts as spheres, they may be projected onto a flat space as shown in \cref{fig:etheneSphereProjection}b where a stereographic projection of the atomic chart in \cref{fig:etheneSphereProjection}a is depicted.
Every point in \TWp\ maps to a G in \TWrho, every trajectory through \TWp\ maps to a zero flux surface in \TWrho, and any closed loop in \TWp\ maps to a volume in \TWrho\ bounded by a zero flux surface and hence characterized by a well-defined energy.
Such volumes are called gradient bundles \cite{morgenstern2015}, and they describe all previously noted zero flux surface-bounded volumes, \eg\ the atomic basin and the bond bundle.

\section{Computational Methods}
\label{sec:methods}

All chemical simulations were performed with the Amsterdam Density Functional \cite{tevelde2001,fonsecaguerra1998,ADF2017authors} \textit{ab initio} software using the Perdew-Burke-Ernzerhof (PBE) functional \cite{perdew1996} and a triple-zeta with polarization (TZP) all-electron basis set.
Calculation of \TWp\ was performed using the Gradient Bundle Analysis tool of the in house Bondalyzer package (by the Molecular Theory Group at Colorado School of Mines) within the Tecpot 360 visualization software \cite{tecplotinc.2013}.

\section{The Topology of the Gradient Bundle Condensed Charge Density}
\label{sec:condencedRhoTopology}

Maxima in \TWp\ typically map to bond paths in \TWrho, as demonstrated in \cref{fig:etheneSphereProjection}a where the three maxima on the carbon atomic chart coincide with the intersections of the carbon-carbon and two carbon-hydrogen bond paths (black paths) with the sphere.
Just as all Gs terminating at the same maxima in \TWrho\ define the atomic basin as a unique volume, all the gradient paths in \TWp\ (\TWcgp s)---\ie\ defined according to \TWcgrad---terminating at the same maxima define a similarly unique gradient bundle, and hence a unique zero flux surface-bounded volume in \TWrho.

\begin{figure}
	\centering
	\includegraphics[width=\textwidth]{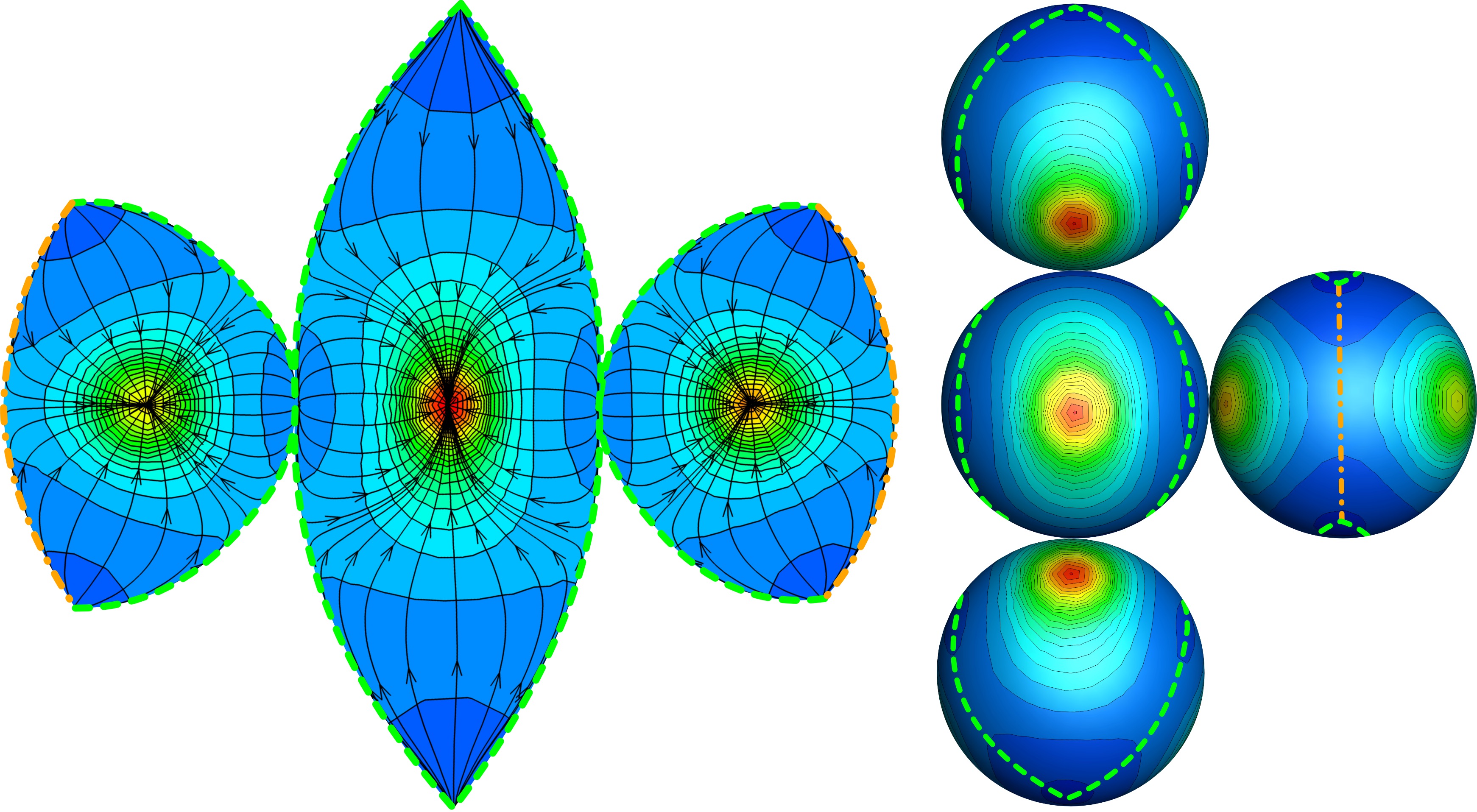}
	\caption{
		\textbf{Left)} Three stereographic projections of \TWp\ for a carbon atom in ethene, centered at the \bd{C}{C} bond path (center) and at each of the \bs{C}{H} bond paths (left and right).
		The three projections together cover the sphere.
		\TWcgp s are shown delineating the three \TWp-basins.
		The \bd{C}{C} and the two \bs{C}{H} basins are demarcated by a dashed green path, and the two \bs{C}{H} basins by a dot-dashed orange path.
		\textbf{Right)} Multiple views of the same \TWp\ mapped onto a sphere.
		The middle left view is centered at the \bd{C}{C} bond path, top/bottom show the same region from above/below, and the right shows the opposite side of the sphere.
		The dashed green and dot-dashed orange paths demarcate the same regions as in the left side of the figure.
		\textit{See electronic version for color images.}
	}
	\label{fig:etheneTripleProjection}
\end{figure}

As an example, shown in \cref{fig:etheneTripleProjection} are stereographic projections for the carbon atomic chart in ethene centered on each of its three maxima.
Also shown are three sets of \TWcgp s (black paths with arrows) each delineating a basin in \TWp.
These basins are bounded by zero flux paths in \TWp\ (dashed and dot-dashed paths), which necessarily map to zero flux surfaces in \TWrho.
The union of these zero flux surfaces partitions \TWrho\ into space-filling regions.
The energy of these regions is well defined and the sum of these energies gives the molecular energy.

These observations invite the following definitions: i) a \textit{bond wedge}\footnote{The term ``bond wedge'' was suggested by Blanco [M.~Blanco, personal communication at the 5\textsuperscript{th} European Charge Density Meeting, 2008].} is the image in \TWrho\ of \TWcgp s terminating at a common maximum in \TWp; ii) a \textit{bond bundle} is the union of two (or more) bond wedges that share an interatomic surface.
As the Bader atom is the union of Gs in \TWrho\ terminating at a common nuclear CP, 
the Bader atom and the bond bundle are conceptually equivalent; one is a basin in \TWrho, and the other a basin in \TWp.

\begin{figure}
	\centering
	\includegraphics[width=0.9\textwidth]{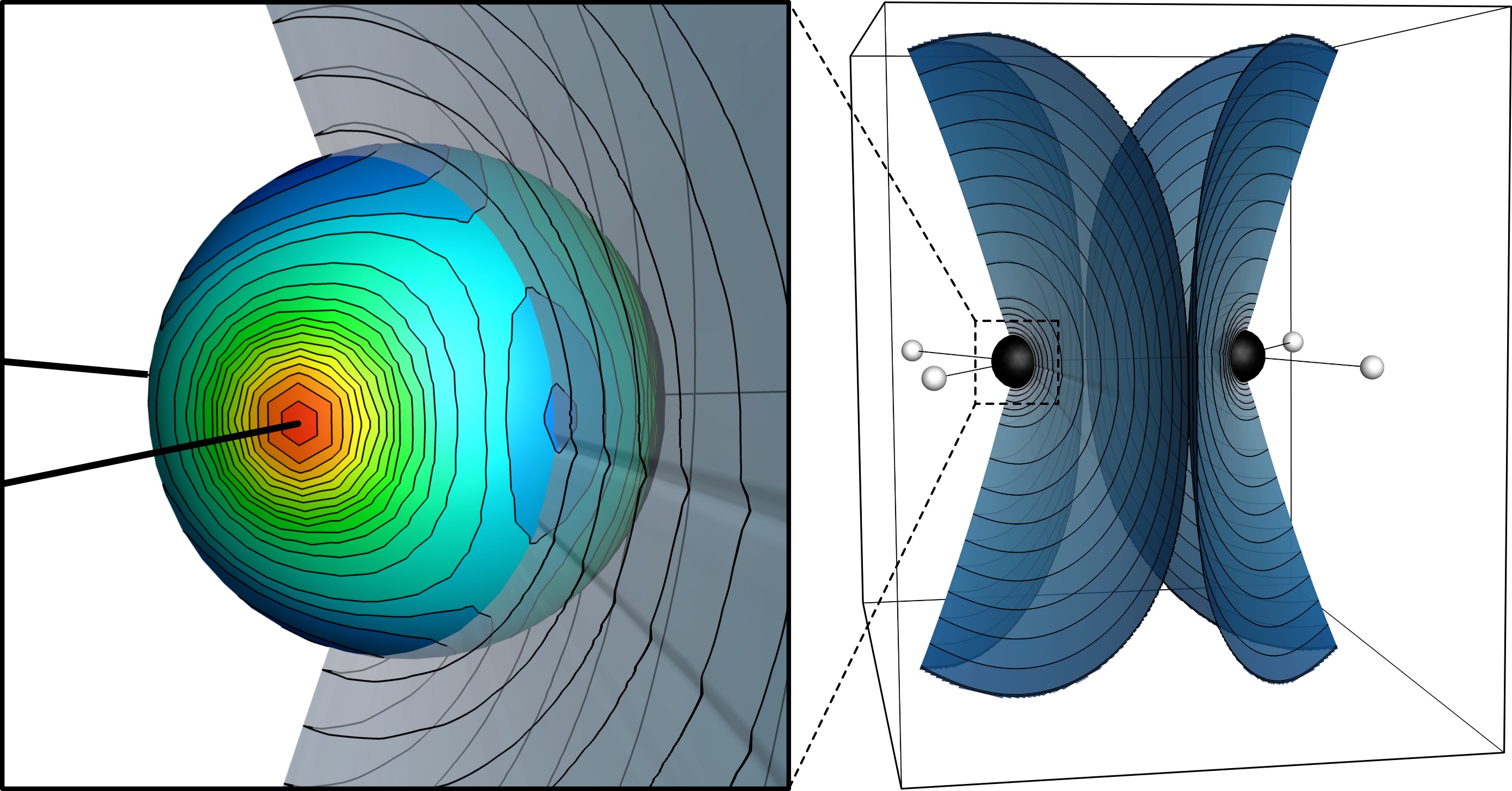}
	\caption{
		\TWp\ for carbon atom in ethene (sphere) shown with \bd{C}{C} bond bundle surfaces (blue) that were identified according to the previously defined special gradient surface criteria.
		Contours on bond bundle surfaces are only shown to enhance perspective.
		\textit{See electronic version for color images.}
	}
	\label{fig:etheneGBABB}
\end{figure}

For the vast majority of organic systems, the bond bundle definition provided here recovers the same regions as those resulting from the earlier definition \cite{jones2009,jones2010}.
This is confirmed in \cref{fig:etheneGBABB}, where the bond bundle surfaces identified according to the earlier definition coincide with those defined by the image in \TWrho\ of the zero flux paths of \TWp\ (the green dashed lines in \cref{fig:etheneTripleProjection}).

\begin{figure}
	\centering
	\includegraphics[width=\textwidth]{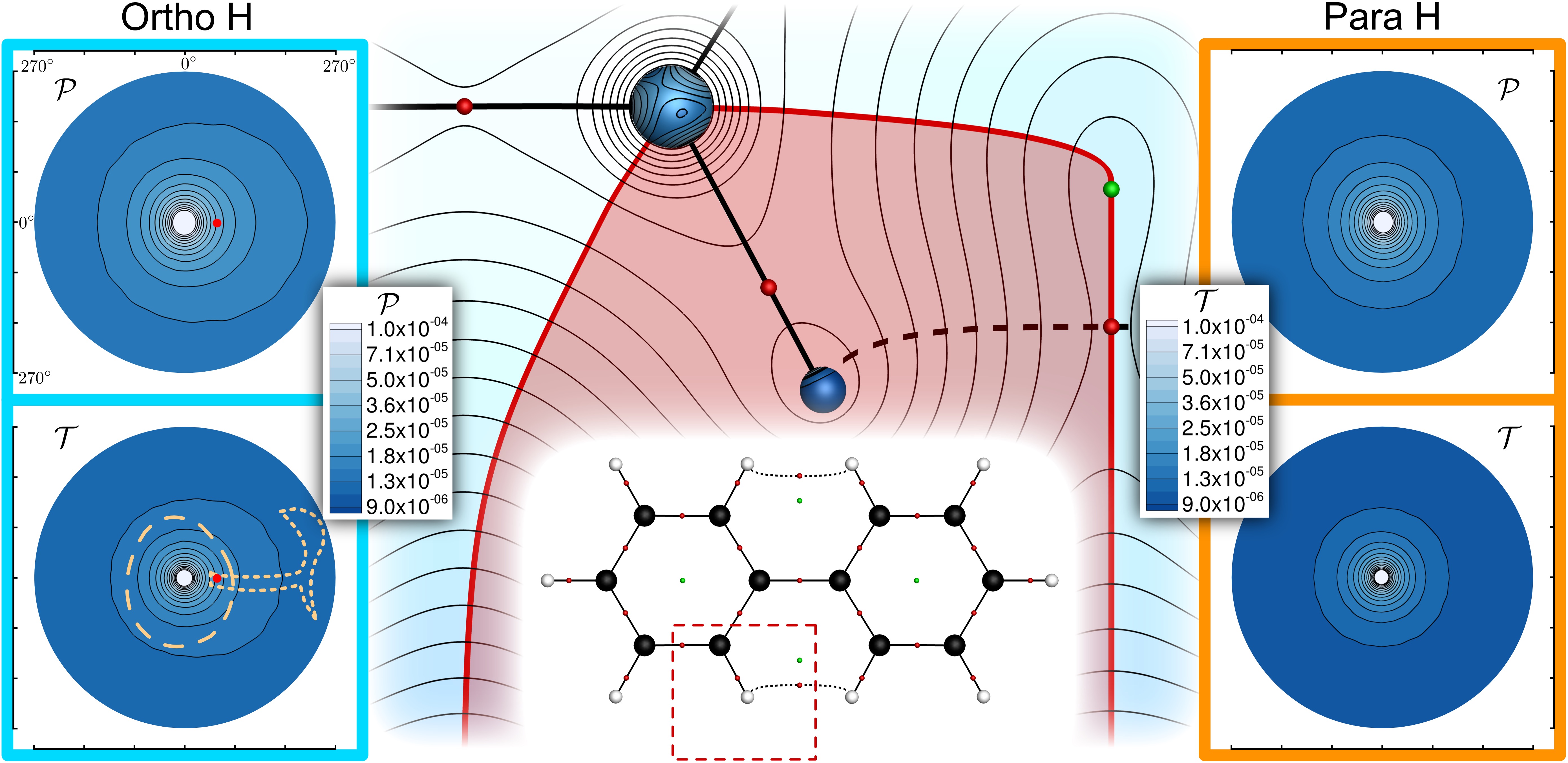}
	\caption{
		Planar biphenyl with stereographic projections of \TWp\ and \TWt\ for the ortho (cyan boxes; left) and para (orange boxes; right) H atoms.
		Center image shows contours of \TWrho\ on the molecular plane (corresponds to the region indicated by a dashed red box in the inset) with GBA spheres for ortho C and H atoms.
		Contours of \TWp\ are mapped onto the spheres.
		The red region shows where the ortho \bs{C}{H} bond bundle intersects the molecular plane.
		Stereographic projections are centered at the \bs{C}{H} bond path with the molecular plane passing horizontally through the projections.
		The intersection of the ortho \bs{H}{H} bond path with the GBA sphere is indicated by a red dot.
		The bottom-left projection includes the boundaries of two gerrymandered districts in \TWt, both of which contain the bond path intersection point.
		\textit{See electronic version for color images.}}
	\label{fig:biphenyl}
\end{figure}

Note that this definition does not require the presence of a bond path, though in organic molecules a maximum in \TWp\ is typically accompanied by a bond path in \TWrho.
There are some noteworthy exceptions, however, as in the case of \bs{H}{H} bonding.
One system in which this occurs is planar biphenyl, where conventional QTAIM analysis reveals bond CPs between ortho hydrogen atoms that are not found in its twisted, lower energy conformation.
The chemical significance of these (and similar) bond CPs is a subject of debate.
Appealing to total molecular and Pauli repulsion energy, molecular orbital analysis, and energy decomposition analysis \cite{poater2006,poater2006a}, antagonists argue that the higher energy of planer biphenyl is primarily a result of destabilizing, steric repulsion between ortho H atoms.
The other side argues that the \bs{H}{H} bond path in fact lowers the energy of the meta-stable configuration, and that ortho H Bader atoms are stabilized relative to the relaxed conformation \cite{matta2003,hernandez-trujillo2007,garcia-ramos2018}.

When analyzed within the new bond bundle perspective, the \bs{H}{H} bond paths of biphenyl (\cref{fig:biphenyl}) do not map to maxima in \TWp.
The bond paths in question lie within the ortho \bs{C}{H} bond bundles.
Obviously the existence of a bond CP and its corresponding bond path is not a sufficient condition for the formation of a bond bundle.
The lack of a \bs{H}{H} bond bundle aside, one could attempt to indirectly scrutinize the energy of the region around the \bs{H}{H} bond path by assessing the energies of the C and H atomic basins or the \bs{C}{H} bond bundles (or their constituent C and H bond wedges).
Being faithful to the QTAIM canons, there are no other unambiguous partitionings that contain the \bs{H}{H} interaction.

We can appeal to the virial theorem \cite{slater1933,bacskay2018} to assess the energy distribution of the H bond wedge.
As Bader has shown, the virial theorem is satisfied within zero flux surface-bounded regions \cite{bader1994}, \eg\ gradient bundles.
Hence, in such regions for stable or meta-stable systems, the total kinetic energy equals minus the total energy.
In addition, the noninteracting kinetic energy accounts for the vast majority of the total kinetic energy \cite{rodriguez2009}.
\cref{fig:biphenyl} includes stereographic projections of the gradient bundle condensed (noninteracting) kinetic energy (\TWt).

Recall that \emph{any} closed loop in \TWp\ corresponds to a zero flux surface-bounded region in \TWrho, and that its energy is well-approximated by \TWt.
For the ortho H atom, two regions have been defined such that the \bs{H}{H} bond path lies within each (dashed orange outlines at the bottom-left of \cref{fig:biphenyl}).
Depending on which region is used, the \bs{H}{H} interactions could be argued to have a smaller or a larger energetic contribution to the system, but this amounts to nothing more than energy gerrymandering.
Forgoing this type of subjective appropriation of energetic significance in discerning chemical phenomena \emph{is fundamental to QTAIM}.

Nothing said here detracts from the relevance of \bs{H}{H} interactions in biological systems, organic crystals, or anywhere else they play a role.
However, if one wishes to perpetuate Bader's well-articulated original vision, then the \emph{unambiguous} zero flux surface-partitioning of the subject region is not optional.

\begin{figure}
	\centering
	\includegraphics[width=0.5\textwidth]{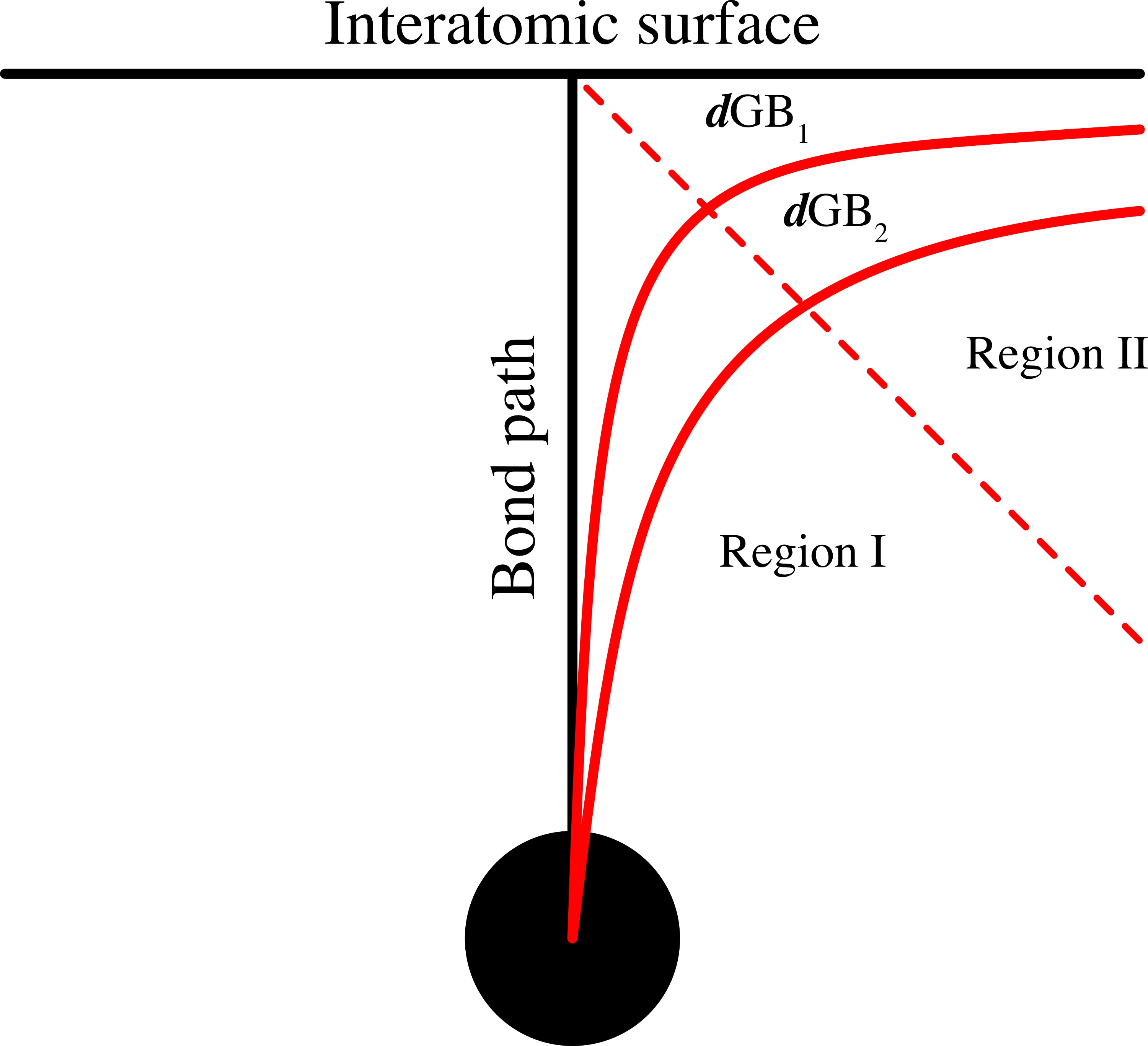}
	\caption{
		Simple representation of the two closest \TWdgb s to a bond path and a line separating the bond path and interatomic surface regions.
	}
	\label{fig:bondConstrants}
\end{figure}

A maximum basin in \TWp\ requires, in addition to a bond path in \TWrho,\footnote{Excepting non-bonding cases, \eg\ lone electron pair regions.} constraints on the relative eigenvalues of the Hessian (\TWhess) along and near the bond path and interatomic surface.
The nature of these constraints can be illustrated by considering a bond path's two nearest neighboring \TWdgb s, one of which (\TWdgb$_1$) contains the bond path, thus coinciding with the interatomic surface, as shown in \cref{fig:bondConstrants}.
If \TWdgb$_1$ is to map to a maximum in \TWp, it must contain more electrons than \TWdgb$_2$, which in turn will be mediated by the relative charge densities in these \TWdgb s over two regions.
First, the region (I) along the bond path where the curvature perpendicular to the bond path is negative.
Over this region the density contained in \TWdgb$_1$ is always greater than the density in \TWdgb$_2$---the greater the negative curvature the more pronounced will be this difference.
In the second region (II), which runs along the interatomic surface, the curvature of the charge density perpendicular to the interatomic surface is positive.
So, in this region the density contained in \TWdgb$_2$ will be greater than that in \TWdgb$_1$.
This difference is minimized by a less curved charge density perpendicular to the interatomic surface.
Combining the two constraints, bond paths will map to maxima in \TWp\ when the curvature perpendicular to the bond path is large and negative and that perpendicular to the interatomic surface is small and positive.

To the extent that \TWhess\ at the bond CP captures the behavior of \TWrho\ over a wider region, a bond bundle will form when the curvature of \TWrho\ at the bond CP is flatter parallel to the bond path and steeper (negatively) perpendicular.
Such behavior will be indicated by a large negative value of the laplacian (\TWlap) at the bond CP.
Bader, arguing from a totally different perspective, came to this conclusion years ago \cite{bader1969} when he asserted that bond energy---or the degree of covalence---was given by \TWlap\ at a bond CP.

\section{Summary}
\label{sec:summary}
It is an established consequence of DFT that chemical phenomena are dictated by \TWrho\ and its redistribution through a physical or chemical process, and though the ability to recover and predict chemical behavior directly from \TWrho\ would be transformative, this goal is as yet unrealized.

QTAIM, as originally formulated, made some progress toward a full charge density representation of chemical phenomena by providing a framework through which to quantify charge redistribution between atoms and to describe the topological changes associated with bond breaking and rearrangement. However, the development of a quantum mechanically rigorous and quantitative description of charge rearrangement that couples easily to the traditional concept of chemical bonding has yet to be achieved.

It seems to us that as we seek the energy mediated traces of chemical behavior, the natural place to look is within a space over which energy is well-defined. The gradient bundle condensed charge density is one such space, and consequently, images in \TWp\ reflect the energetically constrained behavior of \TWrho\ by the full, and only, region to which this behavior corresponds.

The gradient bundle condensed charge density space also allows for a QTAIM-appropriate definition of the bond bundle by revealing the observable bond as the analog of the observable (Bader) atom. For where the Bader atom is an attractor in \TWgrad, the bond bundle is an attractor in \TWcgrad.
Together these observations imply that \TWp\ is an appropriate space in which to describe the energetics of bond breaking and rearrangement.

Investigation to explore the evolution of \TWp\ accompanying physical and chemical processes, across all classes of materials and molecules, will open the door to explanations of chemical phenomena as the manifestation of charge density redistribution within and among observable bonds and atoms.

\section*{Acknowledgments}

Support of this work under ONR Grant No. N00014-10-1-0838 is gratefully acknowledged.

\bibliographystyle{unsrturl}
\bibliography{references}

\end{document}